\documentclass[manuscript]{aastex}

\shorttitle{Low Mass Stars in Praesepe }
\shortauthors{Wang et al.}

\begin{document}

\title{Characterization of the Praesepe Star Cluster by Photometry and Proper Motions
with 2MASS, PPMXL, and Pan-STARRS}


\author{
P.~F.~Wang\altaffilmark{1}, 
W.~P.~Chen\altaffilmark{1,2}, 
C.~C.~Lin\altaffilmark{2}, 
A.~K.~Pandey\altaffilmark{3}, 
C.~K.~Huang\altaffilmark{2}, 
N.~Panwar\altaffilmark{2}, 
C.~H.~Lee\altaffilmark{2}, 
M.~F.~Tsai\altaffilmark{4},
C.-H.~Tang\altaffilmark{4}, 
B.~Goldman\altaffilmark{5},
W.~S.~Burgett\altaffilmark{6}, 
K.~C.~Chambers\altaffilmark{6}, 
P.~W.~Draper\altaffilmark{7}, 
H.~Flewelling\altaffilmark{6}, 
T.~Grav\altaffilmark{8}, 
J.~N.~Heasley\altaffilmark{6}, 
K.~W.~Hodapp\altaffilmark{6}, 
M.~E.~Huber\altaffilmark{6}, 
R.~Jedicke\altaffilmark{6}, 
N.~Kaiser\altaffilmark{6}, 
R.-P.~Kudritzki\altaffilmark{6}, 
G.~A.~Luppino\altaffilmark{6}, 
R.~H.~Lupton\altaffilmark{9}, 
E.~A.~Magnier\altaffilmark{6}, 
N.~Metcalfe\altaffilmark{7}, 
D.~G.~Monet\altaffilmark{10}, 
J.~S.~Morgan\altaffilmark{6}, 
P.~M.~Onaka\altaffilmark{6}, 
P.~A.~Price\altaffilmark{6}, 
C.~W.~Stubbs\altaffilmark{11}, 
W.~Sweeney\altaffilmark{6}, 
J.~L.~Tonry\altaffilmark{6}, 
R.~J.~Wainscoat\altaffilmark{6}, 
C.~Waters\altaffilmark{6}
       }

\altaffiltext{1}{Department of Physics, National Central University, 300 Jhongda Road, Jhongli 32001, Taiwan}
\altaffiltext{2}{Graduate Institute of Astronomy, National Central University, 300 Jhongda Road, Jhongli 32001, Taiwan}
\altaffiltext{3}{Aryabhatta Research Institute of Observational Sciences, Manora Peak, Nainital 263129, India}
\altaffiltext{4}{Department of Computer Science and Information Engineering, National Central University, 300 Jhongda 
                  Road, Jhongli 32001, Taiwan }
\altaffiltext{5}{Max-Planck-Institut f\"{u}r Astronomie, K\"{o}nigstuhl 17, D-69117 Heidelberg, Germany} 
\altaffiltext{6}{Institute for Astronomy, University of Hawai'i, 2680 Woodlawn Drive, Honolulu HI 96822, USA}
\altaffiltext{7}{Department of Physics, Durham University, South Road, Durham DH1 3LE, UK}
\altaffiltext{8}{Department of Physics and Astronomy, Johns Hopkins University, 3400 North Charles Street, Baltimore, MD 21218, USA}
\altaffiltext{9}{Department of Astrophysical Sciences, Princeton University, Princeton, NJ 08544, USA}
\altaffiltext{10}{US Naval Observatory, Flagstaff Station, Flagstaff, AZ 86001, USA}
\altaffiltext{11}{Department of Physics, Harvard University, Cambridge, MA 02138, USA}


\begin{abstract}
Membership identification is the first step to determine the properties of a star cluster.  Low-mass members in 
particular could be used to trace the dynamical history, such as mass segregation, stellar evaporation, or tidal 
stripping, of a star cluster in its Galactic environment.  We identified member candidates 
with stellar masses $\sim$0.11--2.4\,M$_\sun$ of the intermediate-age Praesepe cluster (M\,44), by using Pan-STARRS and 2MASS 
photometry, and PPMXL proper motions.  Within a sky area of 3 deg radius, 1040 candidates are identified, of which 96 
are new inclusions.  Using the same set of selection 
criteria on field stars led to an estimate of a false positive rate 16\%, suggesting 872 of the candidates being true members.
This most complete and reliable membership list allows us to favor the BT-Settl model in comparison with other stellar models.   
The cluster shows a distinct binary track above the main sequence, with a binary frequency of 20--40\%, and a high occurrence 
rate of similar mass pairs.  The mass function is consistent with that of the disk population but shows a deficit of members 
below 0.3 solar masses.  A clear mass segregation is evidenced, with the lowest-mass members in our sample being 
evaporated from this disintegrating cluster.
\end{abstract}



\section{Introduction}

A star cluster manifests itself as a density concentration of comoving stars in space.  Born out of the same molecular 
cloud, the member stars have roughly the same age, similar chemical composition, and are at essentially the same distance 
from us.  Star clusters, therefore, serve as good test beds to study 
stellar formation and evolution.  In order to diagnose the properties of a star cluster, such as its age, distance, size, 
spatial distribution, mass function, etc., it is necessary to identify as completely as possible the member stars.  
In particular, with a sample of members including the lowest mass stars, or even substellar objects, 
one could trace the dynamical history of an open cluster, e.g., the effect of mass segregation, stellar evaporation, 
and tidal stripping in the Galactic environment.

Nearby open clusters are useful in study of their low-mass population.  Praesepe 
(M\,44; NGC\,2632; the Beehive Cluster) is such a rich ($\sim1000$ members) and intermediate-age \citep[757~Myr;][]{gas09} 
stellar aggregation in Cancer, as a member in the Hyades moving group \citep{egg60}, also called 
the Hyades supercluster.  Compared to Praesepe, the Hyades cluster 
itself has a scattered main sequence in the color-magnitude diagram (CMD) because of the 
significant depth with respect to its distance.   The advantages of studying stars in Praesepe are numerous.  First, with 
a distance determination ranging from 170~pc \citep{reg91} to 184~pc \citep{an07}, the cluster is close enough to detect 
low-mass stars or even brown dwarfs.  In this work, we adopted a distance $179\pm2$~pc \citep{gas09}, and 
metallicity [Fe/H]=0.16 \citep{car11}.  Second, the proper motion (PM) of the cluster is distinct from 
that of the field stars, so contamination is minimized when identifying member stars.  
Third, in contrast to a star cluster at birth, for which the spatial distribution of members is governed by the 
parental cloud structure, the stellar distribution in an evolved cluster depends mainly  
on the interaction between members, from which we could investigate the dynamical evolution of the cluster.




Early PM measurements of Praesepe included the pioneering work by \citet{kle27} to identify bright members 
within a 1-deg radius of the cluster center, and by \citet{jon83} who extended the 
detection limit to $V\sim17$~mag to include intermediate-mass members.
\citet{wan95} combined early data and presented a list of nearly 200 PM members. 
Using PMs and photometry, \citet{jon91} identified a list of 
member candidates from $V\sim 9$ to 18~mag within 2\degr\ of the cluster center.  
Using optical and infrared photometry, \citet{wil95} selected member candidates with mass $M > 0.08 M_\sun$ 
and concluded a mass function similar to the field, with no evidence of stellar evaporation.  
\citet{wan11} summarized the photometric surveys of Praesepe members down to the hydrogen-burning limit.
Notably, \citet{ham95a}, with a limiting magnitude of $ R \ga 20$~mag, thereby reaching the stellar mass of 
$\sim0.1$~M$_\sun$, derived a rising mass function toward the low-mass 
end, and presented evidence of mass segregation \citep{ham95b}.  
With the Two Micron All Sky Survey (2MASS) and Digital Sky Survey data covering a sky area of 
100~deg$^2$, \citet{ada02} extended the lower main sequence to 0.1~M$_\sun$, 
and determined the radial density profile of member stars.  \citet{kra07} surveyed a sky 
area of 300~deg$^2$ to identify members by optical to infrared spectral energy distribution,  
and by PM measurements taken from UCAC2 for bright stars or calculated from USNO-B1 and SDSS positions,  
reaching almost into the brown-dwarf regime.  
Their sample of early-type stars is incomplete because of the bright limit of UCAC2, whereas for later-type members 
the incompleteness is caused by the detection limits of USNO-B1 and 2MASS.
Recently \citet{kha13} used SDSS and PPMXL data to characterize the stellar members, including the 
mass segregation effect and binarity.

There have been efforts to identify brown dwarfs in Praesepe.   \citet{pin97} covered one deg$^2$ down to 
$I\sim21$~mag and identified 19 brown-dwarf candidates without spectral confirmation.  \citet{chap05} presented 
deep optical and near-infrared observations covering 2.6~deg$^2$ to a mass limit of 0.06~M$_\sun$.  
\citet{gon06} explored the central 0.6~deg radius region, reaching a limit of $i_{\rm SDSS}\sim24.5$~mag 
corresponding to $\sim$0.05--0.13~M$_\sun$, and identified one substellar candidate.  
\citet{bou10} performed an optical $I_c$ band and near-infrared $J$ and $K_s$ band photometric survey covering 
3.1~deg$^2$ with detection limits of $I_c\sim23.4$~mag and $J\sim20.0$~mag, and found a handful of substellar 
candidates.  The substellar census was augmented by \citet{wan11} who, using very deep optical ($riz$ and $Y$-band) 
photometry of the central 0.59~deg$^{2}$ region of the cluster, identified a few dozen substellar member candidates.  
The first spectroscopically confirmed L dwarf member in Praesepe was secured by \citet{bou13}. 

The stellar mass function of Praesepe was found to rise until 0.1~M$_\sun$ \citep{ham95b, chap05, bak10, bou10}, in contrast 
to the Hyades, which have about the same age but are deficient of very low-mass stars and brown dwarfs.  
Possible explanations include different initial mass functions for the two clusters, or that Praesepe somehow did not experience
as much dynamical perturbation in its environments \citep{bou08}.  A recent study with the UKIRT Infrared Deep Sky
Survey (UKIDSS) Galactic Clusters Survey derived a declining mass function toward lower masses \citep{bou12}.  
One of the aims of this work is to secure a sample of highly probable members to address this issue.  

The spatial distribution of star in a cluster is initially governed by the structure in the parental molecular cloud.  
As a star cluster ages, gravitational scattering by stellar encounters results in mass segregation \citep{spi75}; 
that is, massive stars tend to concentrate toward the center of the cluster, whereas lower mass stars, with a greater velocity 
dispersion, are distributed out to greater radii.  
For Praesepe, \citet{ham95a} combined their observations, complete to $R\sim20.0$~mag and $I\sim18.2$~mag, with those of 
\citet{mer90} with $I\la12$~mag, to show a clear mass segregation effect.  While brown dwarfs may have a 
preferred spatial distribution within a young star cluster \citep{cab08}, they tend to be distributed uniformly 
as the cluster evolves \citet{del00}.
 

Observational attempts to find and characterize members in a star cluster often are sufficiently deep but limited in 
sky coverage, or cover wide areas but are restricted to only brighter (more massive) members.  Studies with large sky 
coverages usually secure membership on the basis of photometry, lacking PM measurements for faint members.  
In this paper, we present photometric (2MASS and Pan-STARRS) and astrometric (PPMXL) diagnostics to select 
the member candidates in Praesepe.  Our sample allows us to characterize the cluster including the binarity, its size, 
the mass function and the segregation effect.  We describe the photometric and PM data in Section~\ref{sec:data}, 
and how we identified probable members in Section~\ref{sec:selection}.  The 
discussion is in Section~\ref{sec:dis}, for which we compare our results with those in the literature.  The binarity is 
discussed, and evidence of mass segregation and tidal stripping is presented.  The paper ends with a short summary as 
Section~\ref{sec:summary}.  

\section{Data Sources \label{sec:data}}

Data used in this study include photometry and PM measurements within a 5-deg radius around the Praesepe center  
(R.A.=$08^{\rm h}40^{\rm m}$, Decl.$=+19\degr 42\arcmin$, J2000).  Archival data were taken from the 
2MASS Point Sources Catalog, PPMXL, and Pan-STARRS.  The 2MASS Point Source Catalog \citep{skr06} has the 
10$\sigma$ detection limits of $J\sim15.8$~mag, $H\sim15.1$~mag, and $K_s\sim14.3$~mag, and saturates 
around 4--5~mag.  The typical astrometric accuracy for the brightest unsaturated sources is about 70--80~mas.
PPMXL is an all-sky merged catalog based on the USNO-B1 and 2MASS positions of 900 million stars 
and galaxies, reaching a limiting $V\sim20$~mag \citep{roe10}.  The typical error is less than 2~milliarcseconds (mas) per 
year for the brightest stars with Tycho-2 \citep{hog00} observations, and is more than 10~mas~yr$^{-1}$ at the faint limit.

Pan-STARRS (the Panoramic Survey Telescope And Rapid Response System) is a wide field (7~deg$^2$) imaging system, with a 1.8~m, f/4.4 
telescope \citep{hod04}, equipped with a 1.4 giga-pixel camera \citep{ton08}.  The prototype (PS1), located atop Haleakala, Maui, USA 
\citep{kai10}, has been patrolling the entire sky north of $-30\degr$ declination since mid-2010.  Repeated observations 
of the same patch of sky with a combination of $g_{\rm P1}$, $r_{\rm P1}$, $i_{\rm P1}$, $z_{\rm P1}$, and $y_{\rm P1}$ 
bands several times a month produce a huge inventory of celestial objects that vary in brightness or in position.  
Deep static sky images and catalog of stars and galaxies are also obtained.  
The PS1 filters differ slightly from those of the SDSS \citep{aba09}.  The $g_{\rm P1}$ filter extends 20~nm redward 
of $g_{\rm SDSS}$ for greater sensitivity and lower systematics for photometric redshift estimates.  SDSS has no 
corresponding $y$ filter \citep{ton12a}.  The limiting magnitudes are $g_{\rm P1}\sim22.5$~mag, $r_{\rm P1}\sim22$~mag, 
$i_{\rm P1}\sim21.5$~mag, $z_{\rm P1}\sim21$~mag, and $y_{\rm P1}\sim19.5$~mag, with the saturation limit of $\sim14$~mag.  
Upon completion of its 3.5 year mission by early 2014, PS1 will provide reliable photometry and astrometry.   
While incremental photometry of PS1 is available at the moment, the calibration of astrometry, hence the PM measurements, 
will need yet to tie down the entire sky, so no PS1 PM data were used here.  The photometric analysis and calibration 
is described in \citet{mag13}.  
PS1 photometry for each detected object has measurements at multiple epochs, but for the work reported here 
only the average magnitude is used.   
In our study, we therefore made use of the 2MASS photometry for stars too bright for PS1, 
plus the PS1 photometry for faint stars, and the PPMXL PMs to select and characterize stellar member candidates.
In matching counterparts in different star catalogues, one arcsecond was used as the coincidence radius 
among PPMXL, PS1, and 2MASS sources.

\section{Candidate Selection \label{sec:selection} } 

Our membership diagnosis relies on grouping in sky position, in PMs, and along the isochrones appropriate 
for the cluster in the infrared and optical CMDs.  
The sources with 2MASS photometric uncertainties greater than 0.05~mag, roughly reaching $J\sim15.2$~mag, $H \sim14.6$~mag, 
and $K_s \sim14.5$~mag, were removed from the sample.  Candidacy was then further winnowed in the $J$ versus $J-K_s$ CMD 
by including only objects with $J-K_s$ colors within 0.3~mag from the Padova isochrones \citep{mar08}.  
This initial, wide range of colors allowed us not to adopt an \textit{a priori} stellar evolutionary 
model, but in turn to put different models to test, as demonstrated below.

\begin{figure}
  \includegraphics[width=0.8\textwidth, angle=270]{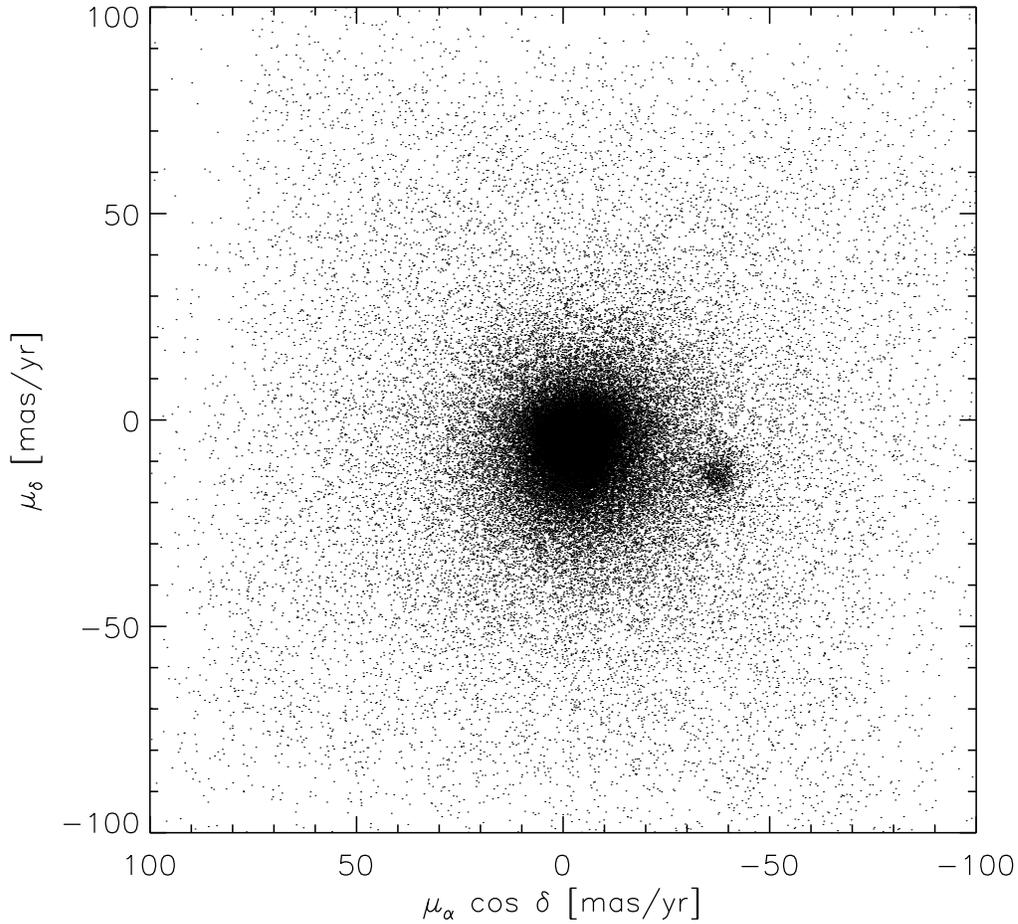}  
  \caption{The PPMXL proper motion vector point diagram of stars toward Praesepe.  Stars within an angular distance of 
            $5\degr$ of the cluster center are analyzed.  Only stars spatially within the 
           central $2\degr$ are displayed here for clarity.   
         }
  \label{fig:ppmxl2deg}
\end{figure}

With the initial photometric sample, we then identified stars with PMs close to that of the cluster.
Obviously the choice of the range is a compromise between the quality 
and the quantity of the candidate list.  The optimal range was decided by how the cluster grouping is blended with the field.   
The PPMXL data toward Praesepe are shown in Figure~\ref{fig:ppmxl2deg}.  
The PM distribution has two peaks, one for the cluster ($\mu_\alpha\, \cos\delta \approx -36.5$\,mas\,yr$^{-1}$, 
$\mu_\delta \approx -13.5$\,mas\,yr$^{-1}$) and the other for field stars ($\mu_\alpha\, \cos\delta \approx -4$\,mas\,yr$^{-1}$, 
$\mu_\delta \approx -3$\,mas\,yr$^{-1}$).  The latter is the reflex Galactic 
motion of the Sun toward this particular line of sight.  The average PM we adopted for the cluster is close to those listed 
by SIMBAD, $\mu_\alpha\, \cos\delta \approx -35.99 \pm 0.14$\,mas\,yr$^{-1}$, and  
$\mu_\delta \approx -12.92 \pm 0.14$\,mas\,yr$^{-1}$ \citep{lok03}.  Naturally, around the peak of the cluster, the distribution is 
dominated by members, and away from the peak the contamination by field stars becomes prominent.   In fact, Praesepe is among a 
few cases where the cluster's motion is clearly separated from that of the field, so the PM distribution exhibits a 
distinct secondary peak due to the cluster.  

\begin{figure}
  \includegraphics[height=0.8\textheight]{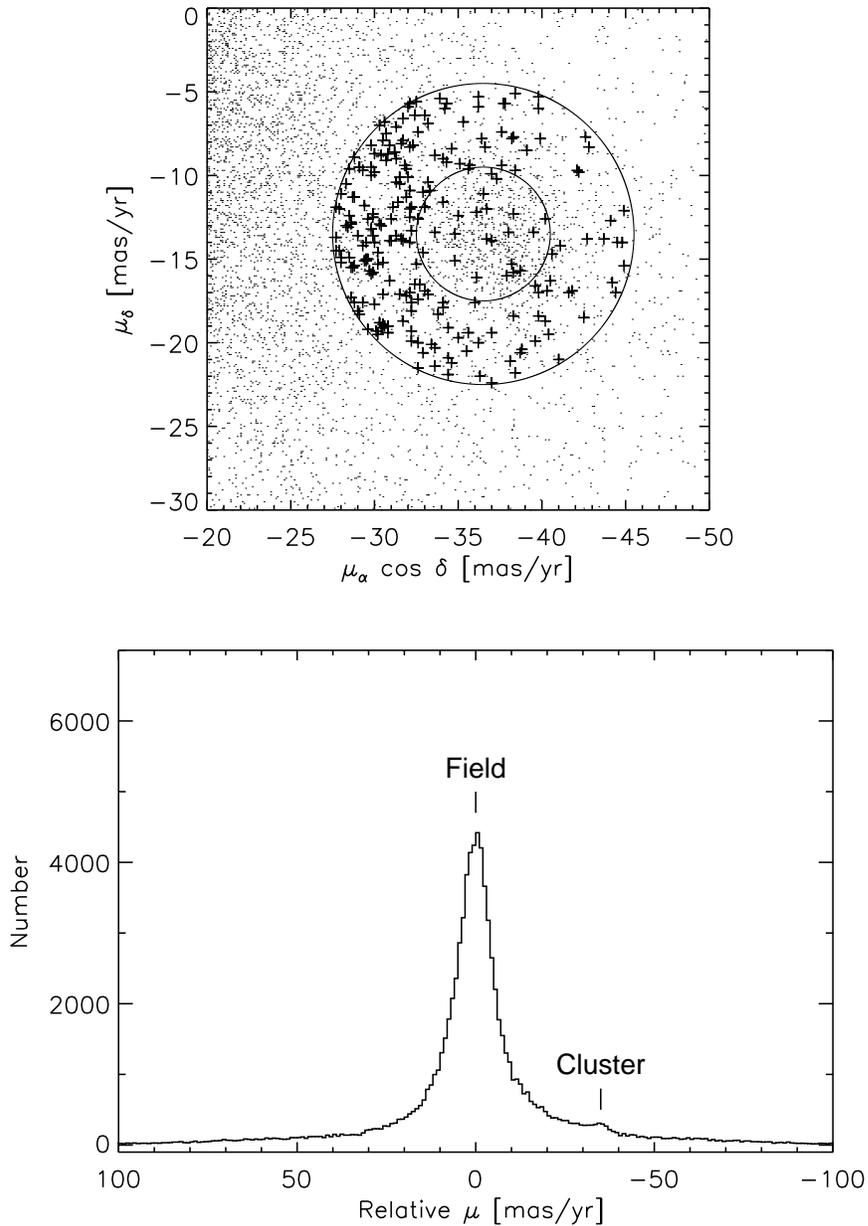} 
  \caption{The 2MASS/PPMXL stars toward Praesepe.  
            Top:~The proper motion distribution. 
                 The two circles illustrate the cases of proper motion range of 
                 $ \Delta \mu =4$~mas~yr$^{-1}$ and of $ \Delta \mu =9$~mas~yr$^{-1}$, respectively.  Stars within 
                 $ \Delta\mu =9$~mas~yr$^{-1}$ but otherwise outside the cluster region (beyond $3\degr$) and photometrically 
                  not following the cluster isochrone, i.e., field stars, are marked with crosses.  
            Bottom:~The projected PM distribution along the line connecting the field centroid and the cluster centroid.  The bump 
                 near $-35$\,mas~yr$^{-1}$ is due to the cluster, which has a standard deviation of 9\,mas~yr$^{-1}$ when 
                 fitted with a Gaussian function.
      }
  \label{fig:pmtrans}
\end{figure}

We exercised two levels of PM selection. First, a Gaussian function was fitted to the secondary (cluster) peak.  
Even through the distribution is known to be non-Gaussian \citep{gir89}, the top part of the peak can be reasonably 
approximated by a Gaussian with a standard deviation of 9\,mas\,yr$^{-1}$.  This is the PM range, namely within  
$\Delta\mu=9$\,mas\,yr$^{-1}$ of the cluster's average PM, that we adopted to select PM membership.  
This range is similar to that used by \citet{kra07} 
(8\,mas\,yr$^{-1}$) or by \citet{bou12} (8\,mas\,yr$^{-1}$ in $\Delta\mu_\alpha\, \cos\delta$ 
and 12\,mas\,yr$^{-1}$ in $\Delta \mu_\delta$).  We note that \citet{bou12} derived, using relative 
PMs on the basis of the UKIDDS data, a different mean motion ($\mu_\alpha\, \cos\delta = -34.17 \pm 2.74$\,mas\,yr$^{-1}$, 
$\mu_\delta = -7.36 \pm 4.17$\,mas\,yr$^{-1}$).  The discrepancy may 
arise because these authors used the median value to choose the center of the PM range, yet the distribution 
is skewed because of the contribution from the field.  The next level of PM selection is $\Delta\mu= 4$\,mas\,yr$^{-1}$, at which 
there is about the same contribution from the cluster and from the field, i.e., a 50\% contamination of the sample.  
Figure~\ref{fig:pmtrans} compares the cases of 4 versus 9\,mas\,yr$^{-1}$.  While bright candidates, including giant stars, 
are not much affected by the choice, the cluster sequence clearly stands out with the narrower PM range even without 
restrictions on position, color, or magnitude.  
The adoption of $ \Delta\mu < 9$\,mas\,yr$^{-1}$ facilitates comparison between our results 
and previous works.  But the $ \Delta\mu  < 4$\,mas\,yr$^{-1}$ sample was still kept for a more reliable selection of candidates.  
Figure~\ref{fig:pmtrans} also shows the PM distribution projected on the line connecting the peak of the field and the peak of the 
cluster.  Even with this projection showing the maximum distinction between the two peaks, the distribution near the cluster 
is overwhelmed by that of the field.    

\begin{figure}
  \includegraphics[width=0.7\textwidth,angle=270]{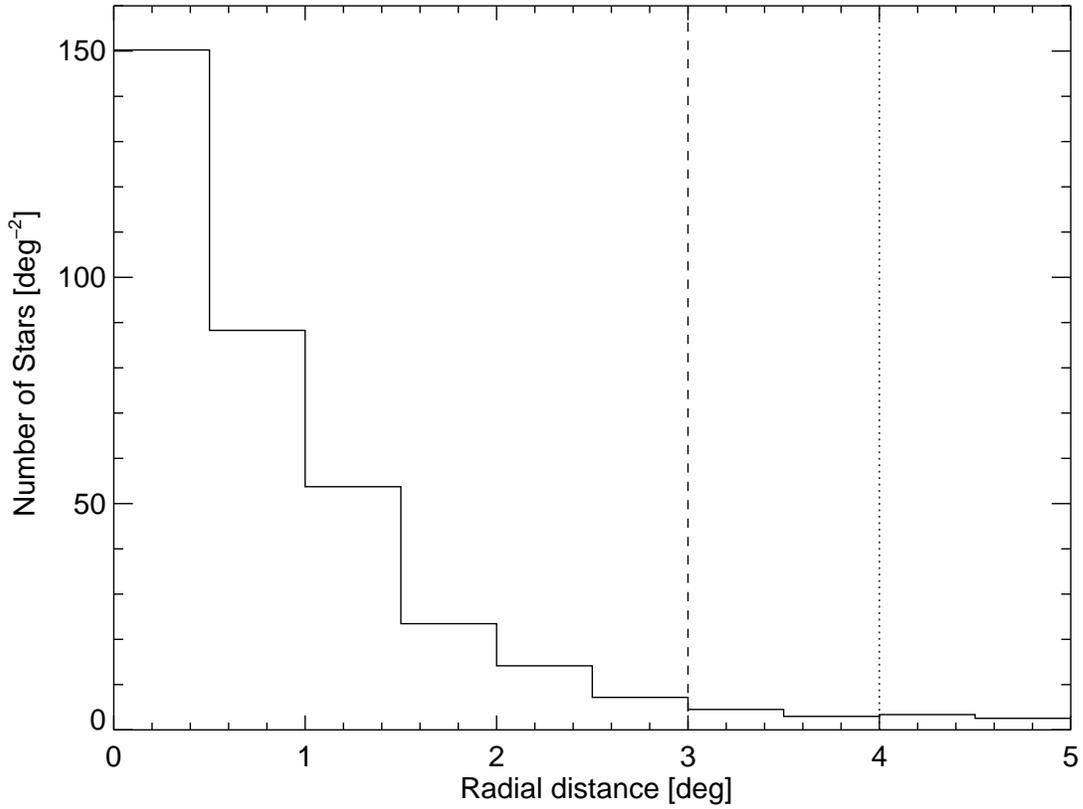} 
  \caption{The radial density distribution of all stars within the entire 5\degr\ field satisfying the isochrone and PM criteria.    
      The vertical line at the 3\degr\ radius marks where we consider the cluster region in our analysis.  The region between radius 
      4\degr\ (shown by another vertical line) and 5\degr\ is used as the field region.
   }
  \label{fig:radden5d}
\end{figure}

Figure~\ref{fig:radden5d} shows the radial density profile of stars following roughly the cluster's isochrone and PM 
within the entire $5\degr$ field.  The surface density decreases monotonically until around $3\degr$, 
then levels off.  Our analysis therefore was conducted within a spatial radius of 3\degr.   At 179~pc, this 
corresponds to a linear dimension of $\sim18$~pc across.  This size is relatively large among the 1657 entries with both angular 
diameter and distance determinations in the open cluster catalog compiled by \citet{dia02}\footnote{Updated to January 2013, 
available at \url{http://www.astro.iag.usp.br/$\sim$wilton/}.}, with the majority having diameters of 2--4~pc.  

\begin{figure}
  \includegraphics[height=0.7\textheight]{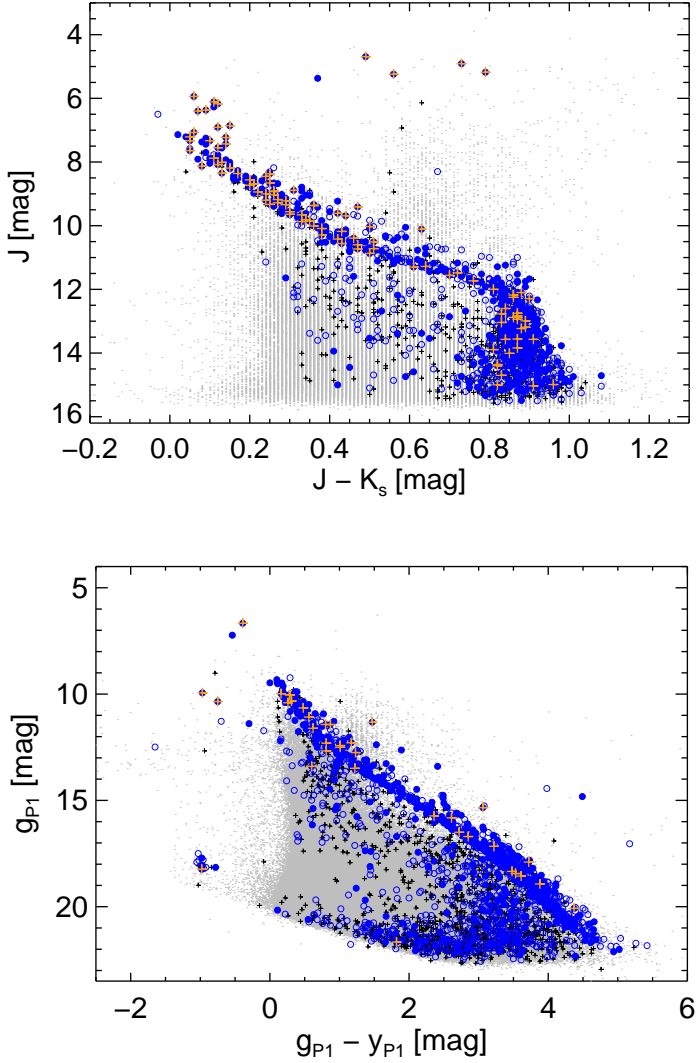} 
  \caption{Top:~The $J$ versus $J-K_s$ CMD for all the stars (gray dots), those with angular distances greater than 
    3\degr\ from the cluster center but with $ \Delta\mu  < 9$~mas~yr$^{-1}$ (small black crosses), those within 3\degr\ from the cluster 
    center and with  $ \Delta\mu  < 9 $~mas~yr$^{-1}$ (blue open circles), and those within 3\degr\ and with 
    $  \Delta\mu  < 4$~mas~yr$^{-1}$ (blue filled circles).  The stars at the very center of the cluster, namely within 30\arcmin, 
    and with $  \Delta\mu   < 4$~mas~yr$^{-1}$ are highly probable members and are marked as orange crosses.  Note 
    the group of blue stragglers beyond the main sequence turn-off point \citep{and98}.   
    Bottom:~The $g_{\rm P1}$ versus $g_{\rm P1} - y_{\rm P1}$ CMD, with the same symbols as in the top panel.  
     The group of stars near $g_{\rm P1} = 18$\,mag, and $g_{\rm P1} - y_{\rm P1} = -1$\,mag include 
     white dwarfs known in the cluster \citep{dob04,dob06}.
   }
  \label{fig:compjjkggy}
\end{figure}

Figure~\ref{fig:compjjkggy} shows the $J$ versus $J-K_s$ and the $g_{\rm P1}$ versus $g_{\rm P1} - y_{\rm P1}$ 
CMDs when the spatial (within or beyond 3\degr\ angular distance from the cluster center) and PM criteria 
(within 9 or 4\,mas\,yr$^{-1}$) are applied.  Even without a preselection by photometry or color, 
the cluster sequence is already evident.  A subsample was chosen with a much restrictive set of parameters, 
namely with the angular distance within the central 30\arcmin, and with $\Delta\mu = 4$~mas~yr$^{-1}$.  This subsample 
is incomplete, but consists of highly secured members, which validates our initial rough selection ranges of magnitude and colors, 
and can be used to compare various stellar atmospheric models.

For the 2MASS/PPMXL sample, photometric 
candidacy is selected in the $J$ versus $J-K_s$ CMD: (i)~for stars brighter than $J\sim12$~mag, from 0.06~mag below to 0.18~mag 
above and perpendicular to the Padova track; for giants there is no photometric restriction, i.e., only the spatial and kinematic criteria 
were applied; (ii)~for fainter stars, from 0.1~mag below to 0.1~mag above and perpendicular to the Siess isochrone.  

For stars fainter than the 2MASS sensitivity, we resorted to the PS1 data collected up to January 2012.  The luminosity function 
toward Praesepe reaches beyond $g_{\rm P1}\sim 21.5$~mag, but our data are limited by the sensitivity of the PPMXL 
dataset at around 21~mag.  To avoid spurious detections, only sources that have been measured more than twice in both $g_{\rm P1}$ and 
$y_{\rm P1}$ bands were included in our analysis.  The $g_{\rm P1}$ magnitudes were derived from the SDSS magnitudes 
\citep[taken from][]{kra07} transformed to the PS1 photometric system \citep{ton12a}, namely, by 
$g_{\rm p1} = g_{\rm SDSS} -0.012 -0.139\, x $, where $x = (g - r)_{\rm SDSS}$.  
For the $y_{\rm P1}$ magnitudes, because SDSS has no corresponding $y$, the transformation from $z_{\rm SDSS}$ was used, 
$y_{\rm P1} = z_{\rm SDSS} + 0.031 -0.095\, x $, where $x$ is again $(g - r)_{\rm SDSS}$.  Because of this, plus the Paschen absorption, 
the transformation to $y_{\rm P1}$ (and to $z_{\rm P1}$) has a larger uncertainty than in other bands \citep{ton12a}.  
In the transformation to either $g_{\rm P1}$ or $y_{\rm P1}$, using the quadratic instead of the linear fit makes little difference.
The bottom panel of Figure~\ref{fig:compjjkggy} plots $g_{\rm P1}$ versus $g_{\rm P1} - y_{\rm P1}$ together with the PS1 
main sequence transformed from \citet{kra07}.  For the PS1/PPMXL sample, the selection range is from 0.15~mag below to 0.4~mag 
above and perpendicular to the \citet{kra07} main sequence transformed to the PS1 system \citep{ton12a}.  

The combination of the 2MASS/PPMXL and the PS1/PPMXL samples contains a total of 1040 stars that satisfy all the criteria 
of photometry (along the isochrone), kinematics (consistent PMs), and spatial (within a 3\degr\ radius) grouping.  
In comparison, there are 168 stars satisfying the identical set of criteria except being with radii between 
4\degr\ and $5\degr$\ (which happens to have the same sky area as the 3\degr\ cluster radius, i.e., $9\pi$\,deg$^2$) 
--- these are 
considered field stars and this number of stars should be subtracted from the cluster region.  So our final list 
contains 1040 member candidates, among which about 872 ($\sim84\%$) should be true cluster members.  Statistically a 
brighter candidate is more likely to be a true member than a fainter candidate because of the field contamination.   
If the stringent criterion of $ \Delta\mu  =4$~mas~yr$^{-1}$ had been used instead, the number of candidates would have 
become 547 within $3\degr$, and 33 between $4\degr$ and $5\degr$, yielding a net of 514 members within $3\degr$, 
yielding a 6\% false positive rate.  

\section{The Updated Member List \label{sec:dis} } 

\begin{figure}
  \includegraphics[height=0.8\textheight]{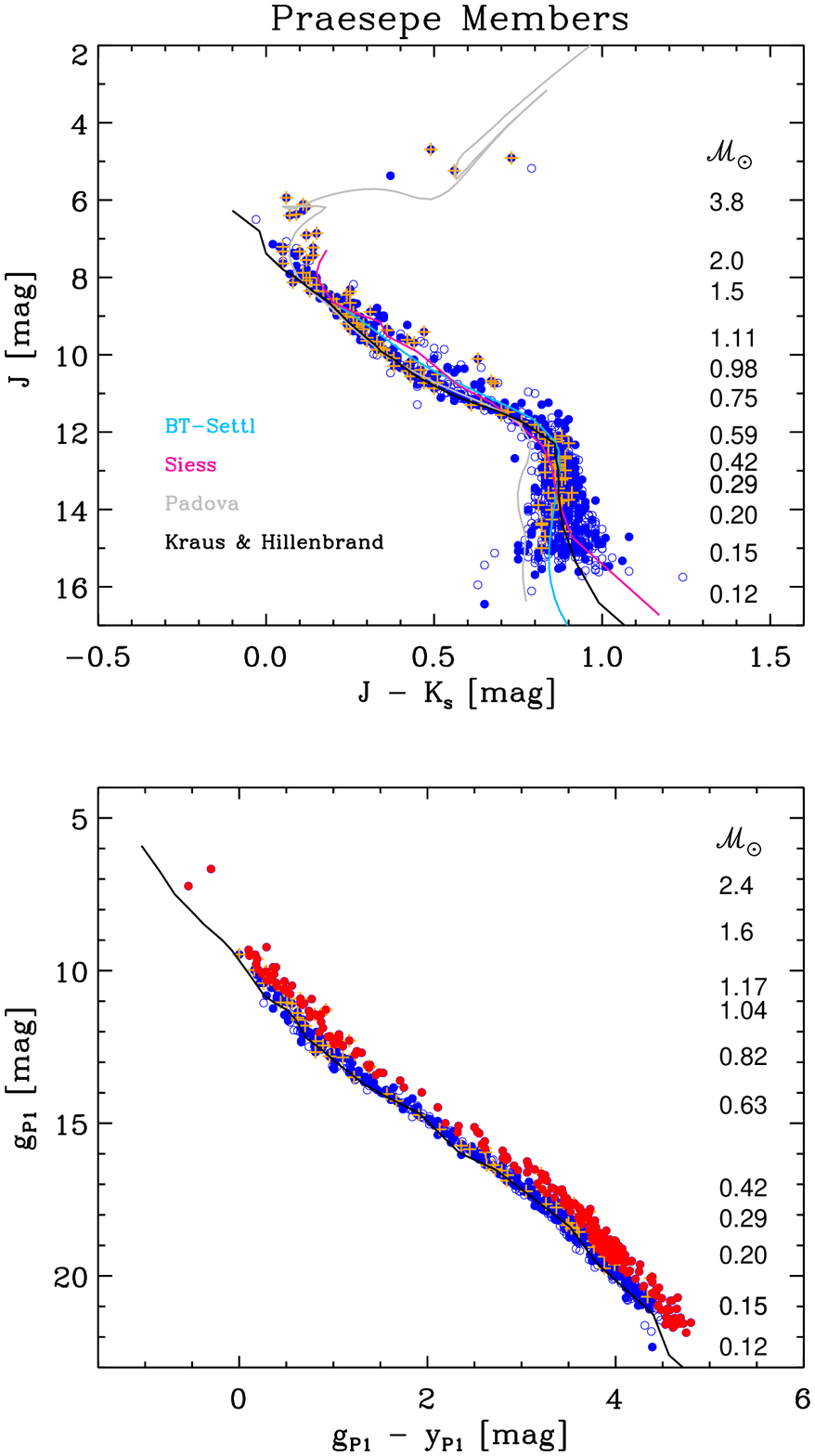} 
  \caption{Member candidates in Praesepe selected on the basis of position, proper motion, and magnitude/color.  
   Top:~The $J$ versus $J-K_s$ CMD, together with the stellar models of BT-Settl \citep{all13,all14}, Siess, Padova, and \citet{kra07}.  
       Selected stellar mass values are labeled.  Symbols are the same as in Figure~\ref{fig:compjjkggy}.  
   Bottom:~The $g_{\rm P1}$ versus $g_{\rm P1} - y_{\rm P1}$ CMD for candidates.  The solid curve is the main sequence from \citet{kra07} 
      transformed to the PS1 system.  Red symbols mark possible binaries. 
   }
  \label{fig:memjjkggy}
\end{figure}

Table~\ref{tab:members} lists the properties of the 1040 candidates.  The first two columns, (1) and (2), are the 
identification number and coordinates.  Columns (3) and (4) give the PM measurements 
and errors in right ascension and in declination taken from the PPMXL catalog.   Subsequent columns, from (5) to (12), 
list the photometric magnitudes and corresponding errors of PS1 $g_{\rm P1}$, $r_{\rm P1}$, 
$i_{\rm P1}$, $z_{\rm P1}$, and $y_{\rm P1}$, and 2MASS $J$, $H$, and $K_s$.  
The (13) column flags if the candidate is possibly binary.  The last (14) column lists the common star name, if any.  
The 2MASS and PS1 CMDs of the members listed in Table~\ref{tab:members} are displayed in Figure~\ref{fig:memjjkggy}, 
along with a selected stellar models:~BT-Settl \citep{all13,all14}\footnote{\url{http://perso.ens-lyon.fr/france.allard/}, 
the latest of NextGen models by \citet{hau99} using the solar abundance of \citet{asp09}}, \citet{sie00}, Padova \citep{mar08}, 
and \citet{kra07}.  To convert the effective 
temperature in the \citet{sie00} models to $J$, $H$, and $K_s$ magnitudes, we made use of the table presented in \citet{ken95}.  
While all isochrones follow roughly each other for $J \la 12$~mag, they differ noticeably toward faint magnitudes.  
The Padova isochrone is too blue to fit the data.  This cannot be caused by reddening because Praesepe is very nearby, 
so is hardly reddened $E(B-V)=0.027$~mag \citep{tay06}.  
The rest four stellar models, though diverging toward the lowest mass end of our data, fit the data equally well.  
The highly secured list of candidates indicates a better fit with the BT-Settl model.

%


\begin{deluxetable}{rrrr r rrrr rrrr r}
\tablecolumns{14}
\tablewidth{0pt}
\setlength{\tabcolsep}{0.02in}
\tabletypesize{\tiny}
\rotate
\tablecaption{Member Candidates of Praesepe }
\tablehead{ 
    \colhead{No.} &
    \colhead{R.A. Decl. (J2000) } & 
    \colhead{ $\mu_\alpha\, \cos\delta$  } &  
    \colhead{ $\mu_\delta$  } & 
    \colhead{ $g_{\rm P1}$  } & 
    \colhead{ $r_{\rm P1}$  } & 
    \colhead{ $i_{\rm P1}$  } & 
    \colhead{ $z_{\rm P1}$  } & 
    \colhead{ $y_{\rm P1}$  }  & 
    \colhead{ $J$  } & 
    \colhead{ $H$  } & 
    \colhead{ $K_s$  } &
    \colhead{ Flag} &
    \colhead{Remarks} \\
    \colhead{ } &
    \colhead{ [deg]  } & 
    \colhead{ [mas~yr$^{-1}$] }  & 
    \colhead{ [mas~yr$^{-1}$]}  & 
    \colhead{ [mag] } & 
    \colhead{ [mag] } & 
    \colhead{ [mag] } & 
    \colhead{ [mag]  } & 
    \colhead{ [mag] } & 
    \colhead{ [mag] } & 
    \colhead{ [mag] } & 
    \colhead{ [mag] } &
    \colhead{  } &
    \colhead{} \\  
    \colhead{ (1) } & \colhead{ (2) } &  \colhead{ (3) } &  \colhead{ (4)} &  \colhead{ (5) } &  \colhead{ (6) } &  \colhead{ (7) } & 
                      \colhead{ (8) } &  \colhead{ (9) } &  \colhead{ (10)} &  \colhead{ (11) } &  \colhead{ (12) } &  
                      \colhead{ (13) } & 
                      \colhead{ (14)}  \\ 
    }
\startdata
413 & 129.7619871 19.7248670 &	$-34.8 \pm 1.1$ & $-13.6 \pm 1.1$ & $12.120 \pm 0.001$ & $\ldots$           & $\ldots$            & $\ldots$            & $\ldots$             & $8.366   \pm 0.026$ & $8.126 \pm 0.021$  & $8.125 \pm 0.021$    & 0 & BD$+$20\,2140	\\
414 & 129.7620587 19.5325438 &	$-37.5 \pm 4.1$ & $-16.9 \pm 4.1$ & $17.618 \pm 0.005$ & $16.347 \pm 0.002$ & $15.095 \pm 0.600 $ & $14.364 \pm 0.001 $ & $ 14.060 \pm 0.003 $ & $ 12.829 \pm 0.022$ & $12.182 \pm 0.021$ & $11.962 \pm 0.019$ & 1 &	\\
415 & 129.7627808 19.4043081 &	$-38.9 \pm 4.1$	& $-16.2 \pm 4.1$ & $19.373 \pm 0.018$ & $18.124 \pm 0.009$ & $16.549 \pm 0.003 $ & $15.841 \pm 0.002 $	& $ 15.494 \pm 0.003 $ & $ 14.261 \pm 0.027$ & $13.643 \pm 0.027$ & $13.407 \pm 0.035$ & 1 & 	\\
416 & 129.7633143 20.0437781 &	$-44.3 \pm 4.1$ & $-13.7 \pm 4.1$ & $14.975 \pm 0.001$ & $13.827 \pm 0.001$ & $13.489 \pm 0.600 $ & $13.074 \pm 0.001 $	& $ 12.927 \pm 0.001 $ & $ 11.867 \pm 0.023$ & $11.209 \pm 0.021$ & $11.051 \pm 0.020$ & 0 & \\
417 & 129.7651196 19.9997784 &	$-31.5 \pm 1.1$ & $-12.6 \pm 1.3$ & $ \ldots$          &  $\ldots$          & $\ldots$            & $12.844 \pm 0.002 $ & $ 12.690 \pm 0.002 $ & $ 7.860  \pm 0.023$ & $ 7.819 \pm 0.016$ & $7.769  \pm 0.018 $ & 0 & HD\,73430	\\
418 & 129.7663424 20.5672773 &	$-38.0 \pm 4.1$ & $-11.3 \pm 4.1$ & $17.691 \pm 0.006$ & $16.496 \pm 0.003$ & $15.336 \pm 0.600 $ & $14.615 \pm 0.001 $	& $ 14.342 \pm 0.002 $ & $ 13.108 \pm 0.025$ & $12.464 \pm 0.024$ & $12.276 \pm 0.021$ & 1 &	\\
419 & 129.7670607 19.5226714 &	$-37.4 \pm 4.1$ & $-12.3 \pm 4.1$ & $14.274 \pm	0.001$ & $13.482 \pm 0.600$ & $13.076 \pm 0.600 $ & $12.831 \pm 0.600 $ & $ 12.601 \pm 0.001 $ & $ 11.562 \pm 0.022$ & $10.987 \pm 0.019$ & $10.857 \pm	0.016$ & 0 & 	\\
420 & 129.7712342 19.7573463 &	$-36.1 \pm 4.1$ & $-15.6 \pm 4.1$ & $19.064 \pm	0.016$ & $17.807 \pm 0.009$ & $16.395 \pm 0.600 $ & $15.616 \pm 0.001 $	& $15.289  \pm 0.003 $ & $ 14.010 \pm 0.024$ & $13.424 \pm 0.030$ & $13.164 \pm	0.028$ & 1 &		\\
421 & 129.7717692 20.1172023 &	$-35.1 \pm 1.1$ & $-14.3 \pm 1.2$ & $9.489  \pm 0.600$ & $9.354  \pm 0.600$ & $9.347  \pm 0.600 $ & $9.375  \pm 0.600 $	& $9.383   \pm 0.600 $ & $ 8.603  \pm 0.030$ & $8.455  \pm 0.026$ & $ 8.413 \pm	0.027$ & 0 & HD\,73429	\\
422 & 129.7754141 19.6768137 &	$-33.7 \pm 1.2$ & $-13.9 \pm 1.2$ & $7.539  \pm	0.600$ & $7.519  \pm 0.600$ & $7.559  \pm 0.600 $ & $7.573  \pm	0.600 $	& $7.586   \pm 0.600 $ & $ 6.857  \pm 0.026$ & $6.769  \pm 0.023$ & $ 6.708 \pm	0.018$ & 0 & HD\,73449	\\
\hline
\enddata
\label{tab:members}
\tablecomments{Table~\ref{tab:members} is published in its entirety in the electronic edition of the journal. 
A portion is shown here for guidance regarding its form and content.}
\end{deluxetable}

Our member candidates have been selected as grouping in five out of six-dimensional photometric and kinematic parameters, 
less only the radial velocity measurements.  Our list hence is more reliable than using photometry alone, 
and is comprehensive in terms of stellar mass and sky area coverage than currently available.   
Among the 1040 candidates, 214 were selected by the 2MASS/PPMXL sample only, 82 by PS1/PPMXL 
only, and 742 by both.  The reason that PS1/PPMXL does not find more candidates is, other than the limit at the bright 
end, because the faintest candidates are very red, $g_{\rm P1} - K_s \approx7$~mag --- in favor of 2MASS detection --- 
and because the PS1/PPMXL data are limited by the brightness limit of PPMXL.  The situation will improve once 
PS1 produces its own PM measurements.  A total of 890 of our candidates coincide with those by \citet{kra07}, 567 with those 
by \citet{bou12}, and 190 with neither.  Of the latter, 96 candidates have not been identified in either \citet{ham95b}, 
\citet{pin97}, \citet{ada02}, or \citet{bak10}. Some of our candidates missed by \citet{bou12} are located in the UKIDSS survey gap.  

Membership identification by photometry alone, e.g., by \citet{gon06} and \citet{bou10}, is vulnerable to significant 
contamination by field stars, so reliable membership could be secured for bright stars only.  To illustrate this, the entire
PS1/PPMXL 5\degr\ sample contains 320,312 stars.  There would have been 2445 candidates if only the photometric and positional 
criteria were set, but the number reduces drastically to 826 once the additional PM criterion ($\Delta\mu \leq  9$~mas~yr$^{-1}$) 
is imposed.

Our member list includes the two stars recently reported by \citet{qui12}, BD$+$20\,2184 (their Pr\,0201=NGC\,2632~KW\,418) 
and 2MASS\,J08421149$+$1916373 (their Pr\,0211=NGC\,2632~KW\,448), to host exoplanets.  
A few candidates found in previous works 
did not pass our PM selection.  For example, stars J083850.6$+$192317 and J084108.0$+$1914901, 
listed by \citet{gon06} as members on the basis of optical and infrared photometry, have PMs 
($\mu_\alpha\, \cos\delta =197.5$~mas~yr$^{-1}$ and $\mu_\delta =79.6$~mas~yr$^{-1}$ for J083850.6$+$192317, and 
$\mu_\alpha\, \cos\delta=-58.4$~mas~yr$^{-1}$ and $\mu_\delta =24.9$~mas~yr$^{-1}$ for J084108.0$+$1914901) 
inconsistent with being part of Praesepe.  Another highly probable member suggested by \citet{gon06}, J084039.3$+$192840, 
already refuted by \citet{bou10} because of its ($I_c - K_s$) color, is indeed not in our candidate list.   
Of the six brown dwarf candidates proposed by \citet[][their Table~5]{bou10}, only three are found in our data, though the 
identification for either stars No.~099, or No.~909 is uncertain because of a nearby star in each case (see the finding 
charts in their Fig.~8).  Only star No.~910 may have a PPMXL counterpart within 10\arcsec, but it has a proper motion  
($\mu_\alpha \cos\delta =-10.5 \pm 7.3$\,mas\,yr$^{-1}$, $\mu_\delta = -10.7 \pm 7.3$\,mas\,yr${-1}$) inconsistent with 
membership.   
The brown dwarf candidate found by \citet{mag98}, NGC\,2632 Roque Praesepe~1, was not in our list because of 
its faint magnitude ($J=21.0$~mag).

\citet{van09} identified, but not tabulated, 24 {\it Hipparcos} members in Praesepe.  With the identifications kindly 
provided by van Leeuwen, we confirm that they are all enlisted in our candidate sample.  
The blue stragglers in the cluster suggested by \citet{and98}, HD\,73666, HD\,73819, HD\,73618, HD\,73210, too bright for PS1, 
are all confirmed to be PM members.  
Our photometric selection precludes the white dwarfs known in the cluster \citep{dob04,dob06}.  They are too faint 
for 2MASS but have been recovered by PPMXL and PS1, illustrated in Figure~\ref{fig:compjjkggy}.  One additional white dwarf
candidate is identified in our data ($\alpha=127.166145\degr$, $\delta=+19.728674\degr$, J2000; 
$\mu_\alpha = -40.4 \pm 5.2$~mas\,yr$^{-1}$, $\mu_\delta = -20.4 \pm 5.2$~mas\,yr$^{-1}$) with $g_{\rm P1} = 18.15$~mag, 
and $y_{\rm P1} = 19.07$~mag.  The white dwarf members follow the general cooling sequence from brighter/bluer to fainter/redder 
in the CMD.  Scaled with white dwarfs in the field, studied by \citet{ton12b} also with PS1 data, the ones in Praesepe have a
cooling time scale of 0.2--0.4~Gyr.

\subsection{Binary Fraction} 

A binary system with identical component stars would have the brightness of either star overestimated 
by 0.75~mag.  A binary sequence therefore is often seen as a swath up to 0.7--0.8~mag above the main sequence 
of a star cluster in a CMD. 
Multiple systems may have even larger magnitude differences.  
\citet{ste95} and \citet{hod99} estimated a multiplicity of $\sim0.5$ for low-mass members in Praesepe.
In both the 2MASS and PS1 CMDs (see Figure~\ref{fig:memjjkggy}), the binary sequence stands out clearly.  Such a distinct binary 
sequence was already noticed by \citet{kra07}.  Note that the $J$ versus $J-K_s$ main sequence is characterized by a 
slanted upper part and turns nearly vertically below the mass of $\sim0.6$~M$_\sun$.   While the upper main sequence allows 
us to gauge the distance (shifting vertically), the vertical segment provides a convenient tool to estimate 
the reddening of a cluster (shifting horizontally).  This fact, however, also means the $J$ versus $J-K_s$ CMD 
cannot be used to evaluate the binarity at the lower main sequence.  Instead, the PS1 CMD shows a monotonic track, so 
is useful for this purpose.  

There is no clear dividing line above the main sequence to separate binaries from single stars.  The bottom panel of 
Figure~\ref{fig:memjjkggy} demonstrates a magnitude difference of 0.5~mag above the main sequence as the dividing line.  
In this case, there are 242 stars above the line, or a binary fraction of about 23\% of the total 1040 member candidates.  
No attempt was made to estimate separately the binarity of the 872 true member versus the 168 interloper samples.  
If the difference is lower to 
0.4~mag or 0.3~mag, the number increases, respectively, to 302 (29\%) or 389 (37\%).  The relatively small increase in 
the binary fraction is the consequence of a distinct binary sequence of this cluster; that is, the binaries in 
Praesepe tend to be of similar-mass systems, as noted, for example, by \citet{pin03}. 
Praesepe also seems to teem with multiple systems, as concluded by \citet{kha13}. \citet{bou12} conducted an 
elaborative analysis on the binarity.  Adopting a brightness range from 0.376 to 1.5~mag above the (single star) main sequence, 
these authors derived a binary frequency of $23.3\pm7.3\%$ for the mass range of 0.45 to 0.2~M$_\sun$, 
$19.6\pm3.8\%$ for 0.2 to 0.1~M$_\sun$, and $25.8\pm3.7\%$ for 0.1 to 0.07~M$_\sun$.  Given the 
uncertainties in membership and binarity assignments, our data do not justify division of the sample into different mass bins, 
and we infer an overall binary frequency (or multiplicity) of at least 20--40\%.



\subsection{Cluster Mass Function \label{sec:mass}}


The stellar mass was interpolated via a least-square polynomial fitting to the $J$ (if too bright in PS1) 
or $g_{\rm P1}$ magnitude using the compilation of \citet{kra07} (their Table~5), and adopting a distance modulus of 6.26~mag. 
The $g_{\rm P1}$ band observations saturate around $g_{\rm P1}\sim14$~mag, corresponding to $J\sim11.5$~mag in our sample, or 
about 0.6~M$_\sun$. 
The masses of our candidates range from $\sim0.11$~M$_\sun$ to $\sim2.39$~M$_\sun$.

\begin{figure}
  \includegraphics[width=0.6\textwidth,angle=270]{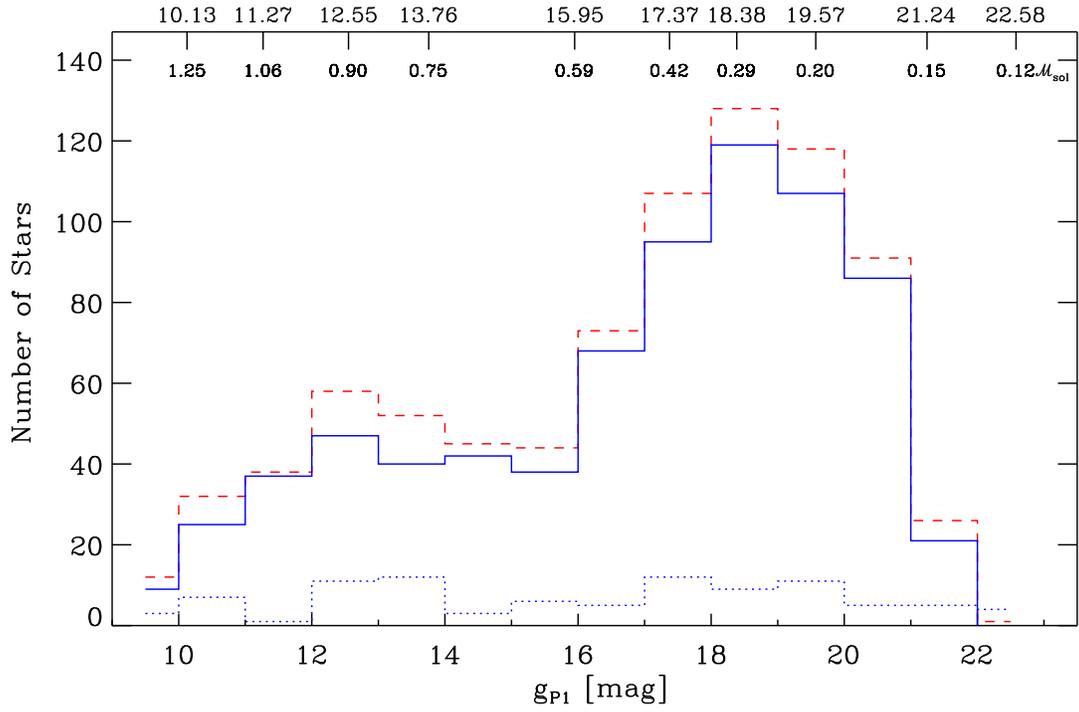}
  \caption{
          The observed $g_{\rm P1}$ luminosity function of member candidates (the red dash line) 
            is subtracted by the field population with the same photometric and PM selection 
                 criteria (blue dotted line) to derive the corrected cluster luminosity function
                 (solid blue line).  The corresponding 
                 stellar mass is labeled at the top in unit of solar mass.
	}
\label{fig:glf}
\end{figure}

The luminosity function of the cluster 
was derived by subtraction of the field contamination.  For field stars, we selected the stars 
satisfying the same PM and isochrone criteria, but with angular distance between $4\degr$ and $5\degr$ from 
the cluster center.  In Figure~\ref{fig:glf}, the $g_{\rm P1}$ luminosity function of the member candidates 
listed in Table~\ref{tab:members} is subtracted by that of the field.  The field distribution is flat, 
as expected, and contributes only as a small correction to the observed luminosity function. 
The corrected luminosity function rises spuriously near the PS1 saturation limit 
of $g_{\rm P1}\sim$11--15~mag, and then turns around near $g_{\rm P1}\sim18$~mag, or mass 
$\sim0.3$~M$_\sun$.  

\begin{figure}
  \includegraphics[width=0.8\textwidth]{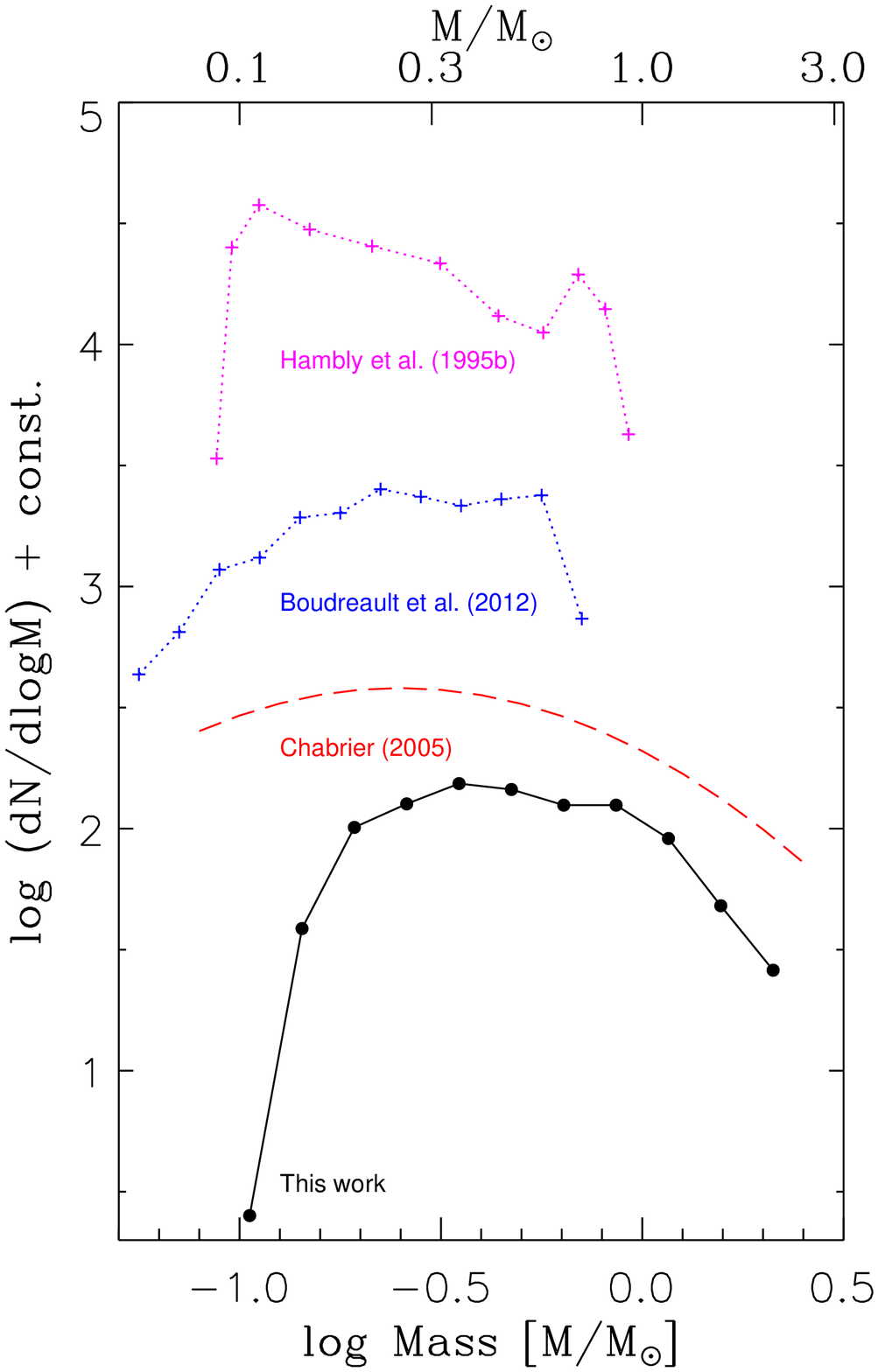}
  \caption{The mass function of Praesepe (solid line).  Also shown are that by \citet{chab05} for the disk population  
     (long-dashed line), and those by \citet{ham95b} (a representative rising mass function) and \citet{bou12} 
     (representing a falling mass function) for Praesepe (dashed lines), each shifted 
      vertically for display clarity.  
	}
\label{fig:mf}
\end{figure}

The mass function of Praesepe members 
is shown in Figure~\ref{fig:mf}.  
We note that this is the mass function for the stellar systems, i.e., with no binary correction.  
Using optical $I_c$ band and near-infrared $J$ and $K_s$ photometric data, 
\citet{bou10} reported a rising mass function in the range from 0.6~M$_\sun$ to 0.1~M$_\sun$ then turning over, 
in agreement with previous works, e.g., by \citet{ham95b}.  This increase in number with decreasing mass
was shown by \citet{wan11} to continue into the brown dwarf regime, peaking around $70$~M$_{\rm Jup}$ then decrease until 
about $50$~M$_{\rm Jup}$.  \citet{kra07} and \citet{bak10} also 
derived a rising, but flatter, mass function.  On the other hand, \citet{bou12}, using also the UKIDSS photometry, but adding 
additional proper motion information, obtained an opposite result, namely, a declining mass function between 0.6~M$_\sun$ and 0.1~M$_\sun$, 
different from those by \citet{ham95b}, \citet{chab05}, 
\citet{kra07}, \citet{bak10}, and \citet{bou10}.  Our sample is more complete at the higher mass end 
than that by \citet{bou12}, but otherwise the mass function is consistent with theirs for stellar masses greater than 
around 0.3~M$_\sun$.  Overall, the mass function we obtained resembles that of the disk population \citep{chab05} for the massive part, 
but shows a deficit of the lowest mass population ($\la 0.3$~M$_\sun$).  


\subsection{Spatial Distribution of Members}

Even the youngest star clusters may have elongated shape \citep{che04}, likely a consequence of 
filamentary structure in the parental clouds.
Subsequent encounters among member stars then circularize the core of a  
cluster.  Mass segregation occurs as energy losing massive stars sink to the center, whereas lower-mass 
members gain energies and occupy a larger volume in space.  Some stars may gain sufficient speed so as to 
escape the system.  The lowest mass members are particularly vulnerable 
to such stellar ``evaporation''.  As the cluster evolves, the internal gravitational pull becomes weaker and 
external disturbances, such as differential rotation, or tidal force from passing molecular clouds and from 
the Galactic disk, act together to distort the shape of a cluster and eventually tear it apart.  
The deformation and tidal stripping are effective even for globular clusters \citep{che10}.  

\begin{figure}
  \includegraphics[height=0.8\textheight] {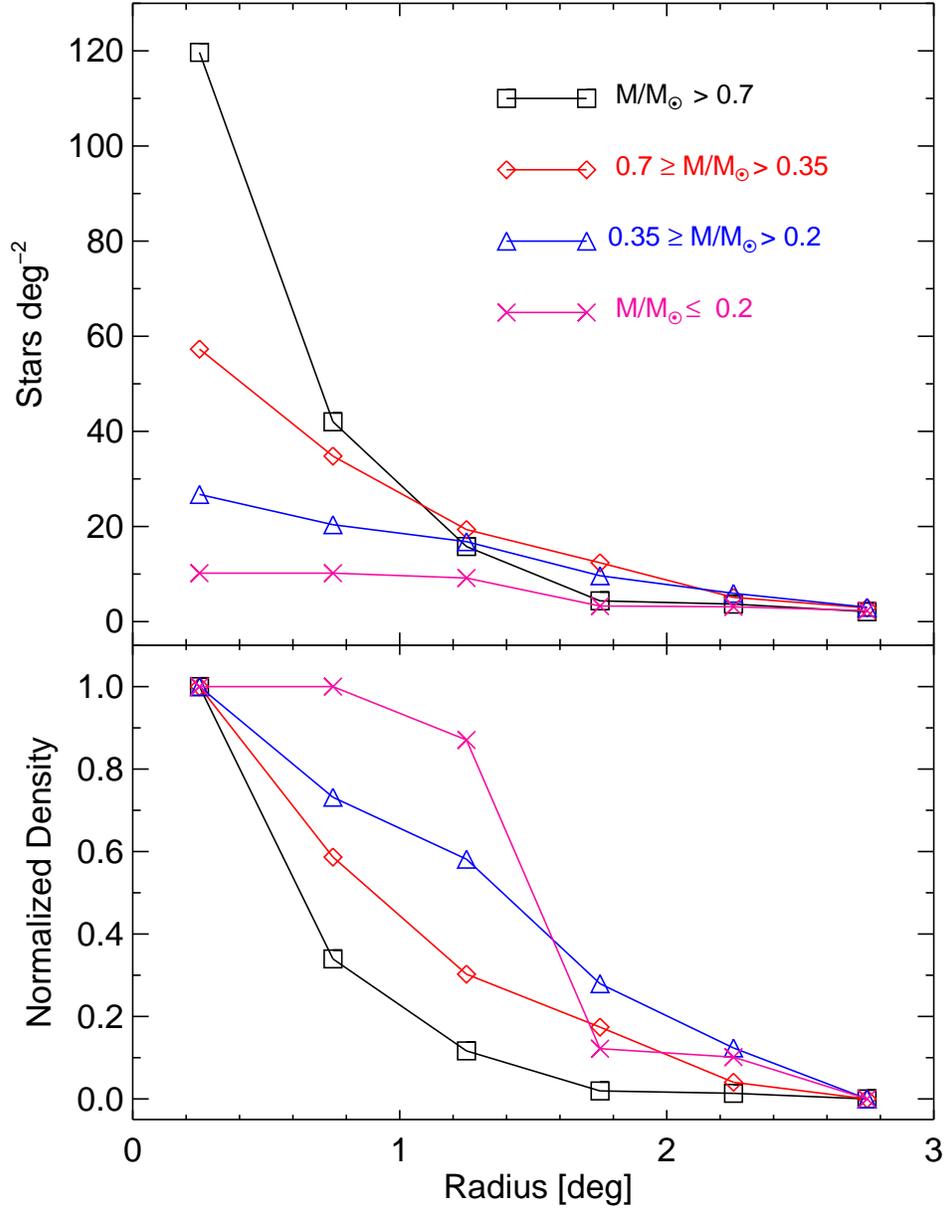}
  \caption{The radial density distribution of the members.  The lines with 
	different colors show different magnitude ranges.   The top panel shows each derived 
        distribution and the bottom panel shows the same but normalized from unity at the center 
        to zero at the edge of the cluster.}
\label{fig:segr} 
\end{figure}

Figure~\ref{fig:segr} shows how the stellar mass correlates with the spatial distribution.  
The radial density profiles have been computed for four different mass groups:~M/M$_\sun \leq 0.2$ (129 stars), 
M/M$_\sun =$0.2--0.35 (256 stars), M/M$_\sun =$0.35--0.7 (332 stars), and M/M$_\sun \geq 0.7$ (323 stars). 
The top panel shows the observed density profiles, while the bottom panel compares the normalized profiles.    
Because of the normalization, no correction of the field contamination is necessary.  
Relatively massive members appear to be centrally concentrated, whereas lower mass members are more scattered spatially, 
a result of mass segregation.  

Mass segregation in Praesepe was well demonstrated already 
by \citet{ham95b}, \citet{kra07}, and \citet{kha13}.  Our result is consistent with that by \citet{ham95b} 
from 0.85~M$_\sun$ to 0.15~M$_\sun$. 
When the radial density distribution shown in Figure~\ref{fig:segr} 
is parameterized with an exponential form, $\sigma(r)~\propto~e^{-\alpha r}$, the least-squared fitting 
yields $\alpha=2.21$ (for members $>0.7~M_\odot$), 0.96 (0.35--0.7~M$_\sun$), and 0.42 (0.2--0.35~M$_\sun$).  
\citet{cab08} suggested that a power-law function may be more appropriate.  In any case, for the faintest sample, 
the density distribution is certainly not exponential.  Instead, it exhibits a sharp truncation beyond 1\degr.  We 
interpret this as a consequence of stellar evaporation.  This further supports the notion of a relative lack 
of low-mass stars in Praesepe, as already demonstrated in Figure~\ref{fig:mf}.  

\begin{figure}
  \includegraphics[height=0.8\textheight,angle=270] {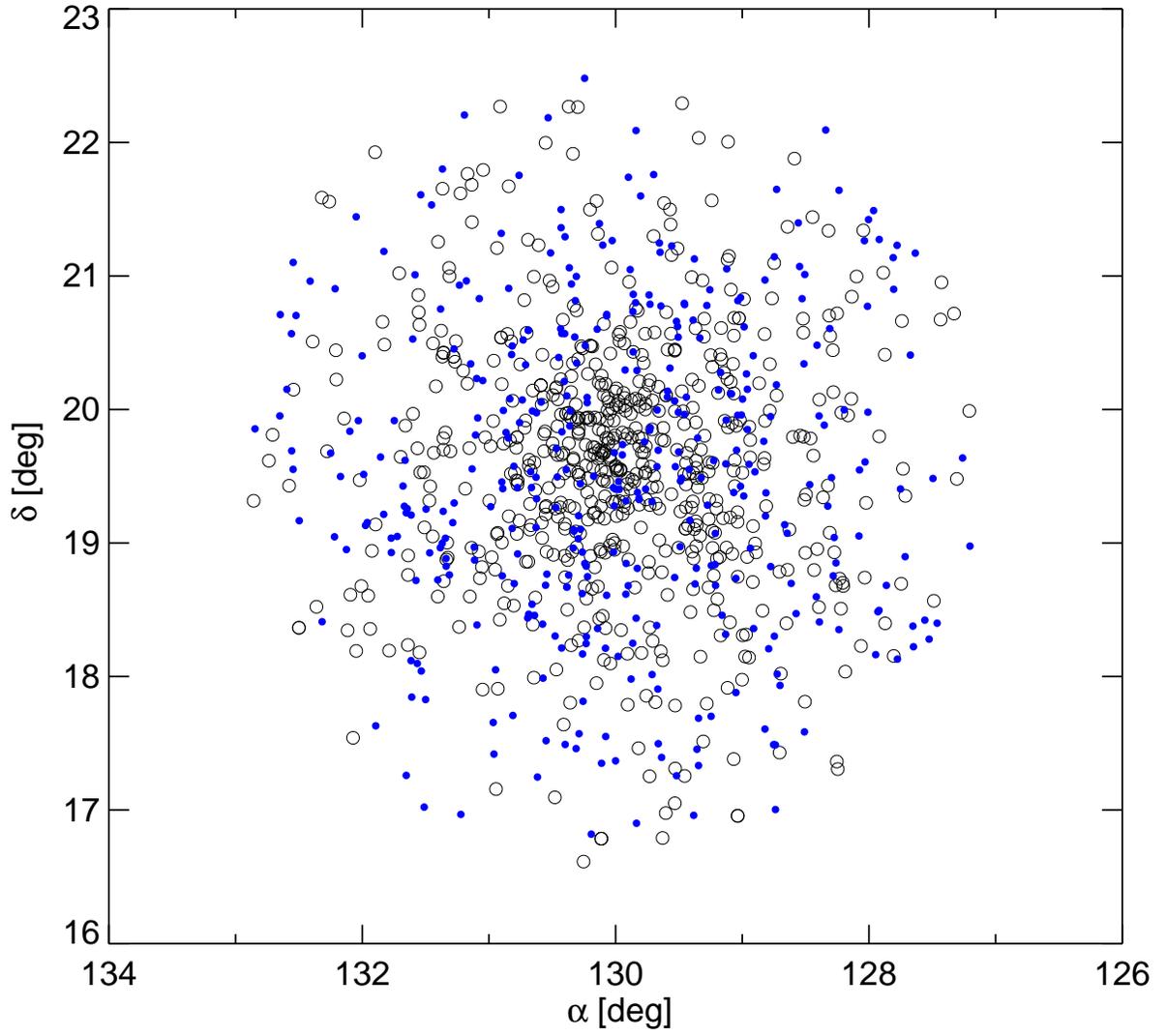}
  \caption{ Positional distributions of stars more massive (open circles) and less massive (solid circles)
             than 0.35~M$_\sun$. 
	}
\label{fig:pos}
\end{figure}

\begin{figure}
  \includegraphics[height=0.8\textheight,angle=270] {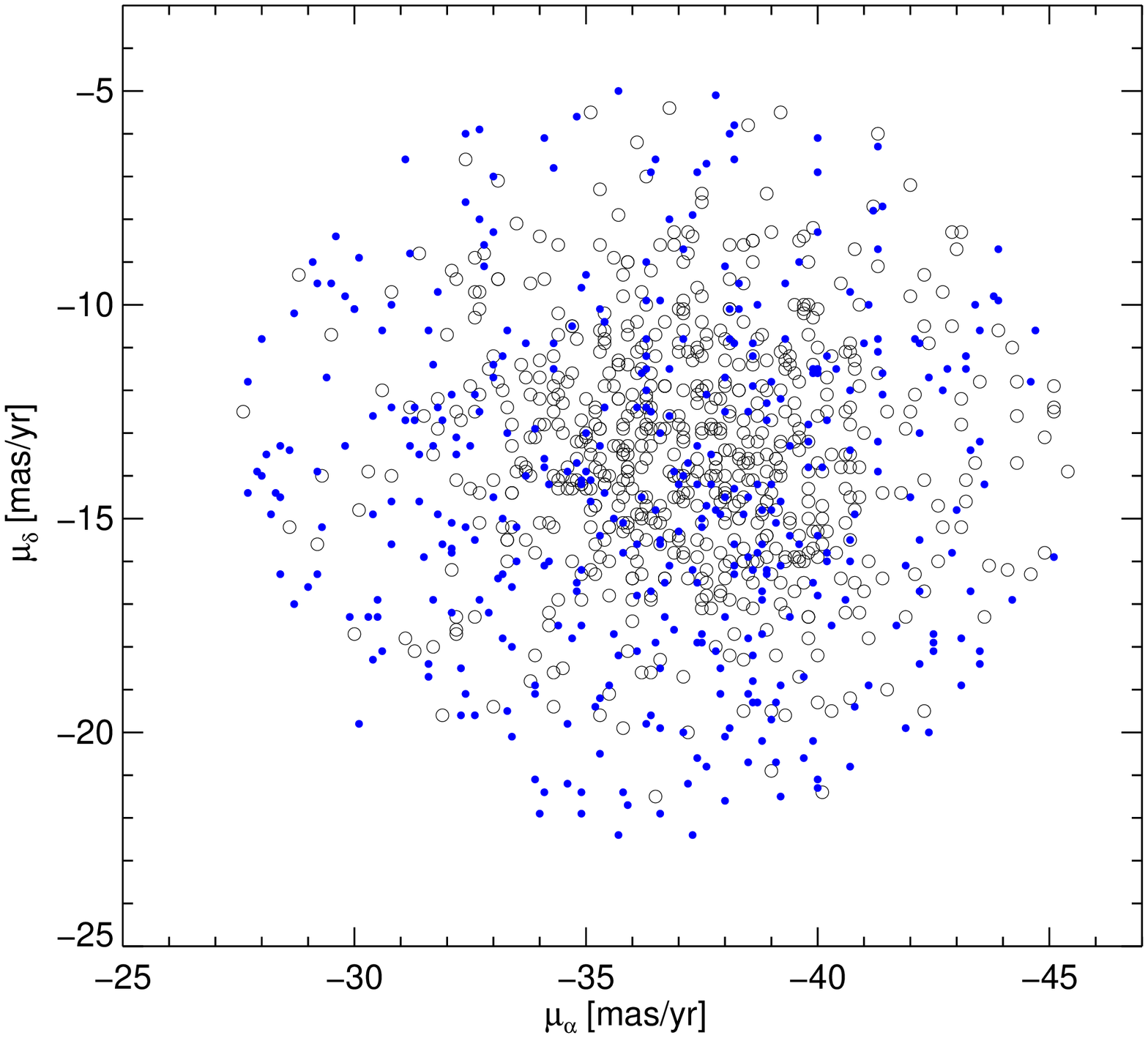}
  \caption{ Proper motion distributions for the same two mass groups of members as shown in Fig.~\ref{fig:pos}.
	}
\label{fig:vel}
\end{figure}

Mass segregation is further manifested by the positional (Figure~\ref{fig:pos}) and PM distributions 
(Figure.~\ref{fig:vel}) of the members; namely, relatively massive members are concentrated in a smaller volume in space, 
and have a smaller velocity dispersion than lower-mass stars.   The average stellar mass in our sample 
is $ \bar{m} \approx 0.59$~M$_\sun$, close to that for a Miller-Scalo initial mass function.  With the 
total number of members  $N=872$, the total stellar mass in the cluster then amounts to at least $\sim520~{\rm M}_\sun$.  
The lowest mass stars, with a declining mass function, do not contribute significantly to the total mass.  With a 
radius $R=9$~pc, the velocity dispersion of the cluster then would be $v \approx (GN \bar{m} /R)^{1/2} = 
0.5$~km~s$^{-1}$, which is noticeably less than the typical value of 1--2~km~s$^{-1}$ for Galactic open clusters.  
At the assumed distance of 179~pc to Praesepe, an intracluster PM dispersion of 1~mas~yr$^{-1}$ corresponds 
to a velocity dispersion of 0.8~km~s$^{-1}$.  Our data thus are not precise enough to measure any PM 
gradient among members.  

The evidence is mounting that Praesepe is dissolving.  It is spatially extended with a sparse stellar density.  
\citet{hol00} suggested that Praesepe might consist of two merging clusters.  
The relatively high fraction of equal mass pairs (and of multiples) may be the consequence of occasional 
stellar ejection during three-body encounters \citep{bin87}, or during the merging process.   
Relevant time scales for a dissolving star cluster include: ($i$)~the dynamical (crossing) time scale,  
$\tau_{\rm dyn} \approx 2R/v$, ($ii$)~the relaxation time, $\tau_{\rm relax} \approx \tau_{\rm dyn} \, 0.1\, N/\ln N$, and
(iii)~the evaporation time, $\tau_{\rm evap} \approx 100\, \tau_{\rm relax}$ \citep{bin87}. 
For Praesepe, these time scales are $\tau_\mathrm{dyn} = 3.6 \times 10^{7}$~yr, 
$\tau_\mathrm{relax} = 4.6 \times 10^{8}$~yr, and $\tau_\mathrm{evap}=4.6\times 10^{10}$~yr, respectively. 
The lowest-mass members, having an average escape probability \citep{spi87} several times of that for the 
most massive stars, are particularly susceptible to ejection.  The Praesepe cluster therefore is 
almost fully relaxed, and tidal stripping has occurred, starting with the lowest mass members being witnessed to 
escape from the cluster.  

\section{Summary} \label{sec:summary}

We have conducted a photometric and proper motion selection of member stars of the Galactic 
open cluster Praesepe, using 2MASS, PPMXL and Pan-STARRS data.  Our sample is comprehensive 
in terms of sky area (3\degr\ radius), limiting magnitude ($g_{\rm P1} \sim21$~mag), and 
reliability ($\sim16\%$ false positive rate).  A total of 1040 member candidates are identified, 
872 of which are highly probable members, down to about 0.1 solar masses.  While for members more 
massive than 0.6~M$_\sun$, the Padova isochrone works well, the BT-Settl atmospheric model fits better
toward fainter magnitudes.  The binary frequency of Praesepe members is about 20--40\%, 
with a relatively high occurrence of similar mass pairs.  The mass function is consistent with that 
of the disk population, but with a deficit of stars less massive than 0.3~M$_\sun$.  Members 
show a clear evidence of mass segregation, with the lowest mass population being evaporated 
from the system.  At the faint magnitude end, the bottleneck of membership 
selection for very faint objects remains the sensitivity of the PM measurements.  
Once the PS1 completes its survey in early 2014, increasing the photometric depth and the stellar 
PM baseline to more than 3.5 years, we expect to secure member lists for nearby star clusters 
well into the substellar regime. 

We thank the referee, Jos\'e A. Caballero, who provided very constructive comments on an earlier version 
to greatly improve the quality of the paper.  We are grateful to Steve Boudreault for providing published 
data to produce Figure~7.  
The Pan-STARRS1 Surveys (PS1) have been made possible through contributions of the Institute for Astronomy, 
the University of Hawaii, the Pan-STARRS Project Office, the Max-Planck Society and its participating institutes, the Max Planck 
Institute for Astronomy, Heidelberg and the Max Planck 
Institute for Extraterrestrial Physics, Garching, The Johns Hopkins University, Durham University, the University of Edinburgh, Queen's University 
Belfast, the Harvard-Smithsonian Center for Astrophysics, the Las Cumbres Observatory Global Telescope Network Incorporated, the 
National Central University of Taiwan, the Space Telescope Science Institute, the National Aeronautics and Space Administration under 
Grant No. NNX08AR22G issued through the Planetary Science Division of the NASA Science Mission Directorate, the National Science Foundation 
under Grant No.~AST-1238877, and the University of Maryland.  The NCU group is financially supported partially by the 
grant NSC101-2628-M-008-002.

%


\end{document}